# Enhancing Communication Quality between Government and Citizens: The Role of AI Modification

Ruiyu Zhang[1] & Lin Nie[2]

## Abstract

This study integrates critical AI scholarship with relational communication theories to explain how AI language modifications shape the quality of government–citizen communication. Distinguishing between informational-cognitive quality (clarity, ease of response) and expressive-constitutive quality (politeness, respectfulness, feeling heard, trust, urgency, empathy), we hypothesize that AI yields uncontested benefits for the former but contested effects for the latter, potentially enhancing relational markers while muting authentic emotional cues. Using a vignette-based survey with 220 citizens and 214 civil servants in China, we assess perceptions across five interaction contexts: service requests, policy inquiries, complaints, suggestions, and emergencies. Results from paired t-tests and mixed-effects regressions support the claim that AI enhances both informational-cognitive and expressive-constitutive quality from the perspectives of citizens and civil servants, with significant improvements in clarity, politeness, satisfaction, trust, and empathy, but provide no consistent evidence of urgency or empathy signals. These findings suggest that concerns over algorithmic emotional flattening may be overstated or context-specific; they offer theoretical insights into AI-mediated public interactions and practical implications for fostering trust and efficiency in digital governance.

This article has been published in ***Information Polity***, 2026. The final version of record is available at: https://doi.org/10.1177/15701255261427986

*Cite*:

Zhang, R., & Nie, L. (2026). Enhancing Communication Quality between Government and Citizens: The Role of AI Modification. *Information Polity*. https://doi.org/10.1177/15701255261427986

[1] Department of Politics and Public Administration, The University of Hong Kong, Pokfulam, Hong Kong, China.

[2] Department of Applied Social Sciences, The Hong Kong Polytechnic University, Hung Hom, Hong Kong, China. Email: lin-apss.nie@polyu.edu.hk

## Introduction

The ongoing digital transformation is reshaping public management, triggering a profound evolution of the relationship between citizens and the government. Since the late 1990s, governments around the world have employed digital tools to enhance citizen engagement (Dunleavy et al., 2006; Torres et al., 2005). The emergence of artificial intelligence (AI) technologies offers innovative solutions for rebuilding government–citizen relations. AI tools, particularly chatbots and natural language processing systems, are increasingly integrated into public administration systems (Wirtz et al., 2019; Yigitcanlar et al., 2024) and have been found to enhance the professionalism and courtesy of public service communications (Ju et al., 2023; Li & Wang, 2024). As governments strive to meet the diverse and complex demands of society, it is crucial to understand AI's capabilities and limitations in facilitating effective communication. Effective government–citizen communication is highly associated with citizen satisfaction and trust (Mahmood et al., 2020; Welch et al., 2005). While previous studies have highlighted AI's potential to improve the interactive experience between government and citizens, they have not comprehensively evaluated its effectiveness across multiple dimensions of communication quality. Moreover, because it is two-way, communication quality should concern both government- and citizen-side perceptions; however, limited empirical evidence provides a holistic understanding of AI's role in influencing both sides simultaneously.

This study evaluates how AI-assisted interactions affect the quality of communication between citizens and the government. Among the broad range of AI functions, we focus on AI language modification, which refers to the use of AI tools—particularly large language models—to improve the clarity, tone, politeness, and emotional expression of written communication (Hancock et al., 2020). Prior research on AI modifications has explored its applications in crisis management (Xiao & Yu, 2025), education (Purcell et al., 2024), workplace communication (Kadoma et al., 2024), and mental health support (Liu et al., 2022). We draw on these foundations and further examine (1) whether AI modifications affect the perceived communication quality by both citizens and civil servants; (2) how AI-modified communication affects citizens' perceptions toward government agencies; and (3) how AI-modified communication assists civil servants in addressing citizens' demands.

To investigate these questions, we focus on written government–citizen communication channels, such as government social media accounts, web-based message platforms, and government mailboxes, which have proliferated across countries (Sivarajah et al., 2015; Stone et al., 2022; Wang et al., 2025). These channels or platforms enable citizens to express their demands and/or grievances to relevant government agencies and expect responses/solutions. Via online media, the government and citizens are indirectly linked through written language. Traditionally, citizen messages and government responses were exchanged without language processing. While unpolished communication can lead to misunderstanding and administrative inefficiency, it may also carry authentic emotional signals that convey urgency or grievance. This study examines whether AI modifications can improve perceived communication quality, while remaining attentive to what may be lost when messages are standardized. We select real-world government–citizen communication scenarios to design a vignette-based survey in China. Findings indicate that AI modifications enhance the clarity, politeness, and satisfaction of written communications between citizens and the government. Specifically, citizens perceive AI-modified governmental responses as more trustworthy and empathetic, while civil servants find AI-modified citizen messages clearer, more respectful, and easier to address. Nevertheless, AI's effectiveness in conveying urgency and emotional empathy remains inconsistent, pointing to important areas for future technological refinement.

The digital trend of public administration makes this study a timely and important research. This study contributes to the theoretical discourse on government–citizen interactions and AI integration, and offers practical insights for public administrators seeking to leverage AI to enhance government communication quality and citizen perceptions. The remainder of the article is organized as follows. The next section reviews existing scholarship about government–citizen interaction and AI's integration in the interactive process, followed by an analytical framework and hypotheses development. In the Methodology section, the data collection procedure and research design are detailed. The Results section presents the empirical findings. Lastly, research conclusions, limitations, and practical implications are discussed.

## Literature Review

### *Government–Citizen Interaction and Communication*

Government–citizen interaction encompasses a wide range of relational processes, including feedback, collaboration, and co-production in public service and policymaking (Aichholzer & Strauß, 2016; Bokayev et al., 2023). Communication is a foundational component of interaction, referring to the exchange of information between governments and citizens. Historically, public communication emphasized one-way dissemination of information, often through official announcements, notices, or structured consultations (Mergel, 2013; Thomas & Streib, 2003). These mechanisms—while providing transparency—offered limited channels for reciprocal interaction or participatory decision-making. Even as digital tools gained traction, studies suggest that many virtual interactions have continued to mirror traditional top-down structures (Brainard & McNutt, 2010). Public communication remained informational in nature, with only limited transactional and collaborative exchanges, resembling the old public administration patterns. Nonetheless, the emergence of e-government marked a turning point, enabling governments to shift from static, unidirectional platforms to more dynamic systems that facilitate two-way communication. Although the early stages of e-government adoption were somewhat chaotic and fragmented, they prompted the development of structured growth models to guide institutional transformation (Layne & Lee, 2001). As municipalities increasingly embrace digital infrastructure and e-government, modern governance is gradually characterized by networked coordination, citizen-centric services, and external collaboration (Ho, 2002). These shifts laid the groundwork for deeper reconfigurations of government–citizen relations, pushing beyond communication as mere transmission toward interaction as ongoing engagement.

Recent work suggests that while digital channels have become more widely used, citizens still prefer traditional modes for certain high-stakes or complex services, indicating the importance of maintaining diverse, task-appropriate channel strategies (Pieterson & Ebbers, 2020). Online systems can also magnify inequalities in engagement. For instance, digital tools have encouraged more frequent contact among some groups, but others remain less likely to initiate interaction. This phenomenon is called the digital divide (McNeal et al., 2008). That being said, well-designed e-government platforms have been found to improve perceived service quality and citizens' trust in government (Amosun et al., 2022). In parallel with these functional developments, scholars have drawn attention to the need for public institutions to embed relational capacities, such as empathy, into the design and practice of digital tools. Empathy is increasingly seen as a skill that can be cultivated within administrative contexts to foster connection, responsiveness, and a more human-centered public encounter (Edlins, 2021). Recent literature further emphasizes that empathy should be operationalized as both a felt experience and a communicative act, with important implications for civil servants managing conflicts and resource allocation (Mussagulova, 2026). Together, these findings point toward a vision of government–citizen interaction—one that blends technical efficiency with emotional and experiential quality.

Recent advancements in digital technologies have opened new avenues for more inclusive and far-reaching engagement, but they also highlight significant complexities. Digital tools potentially expand service accessibility, transparency, and public participation (Tejedo-Romero et al., 2022). Yet many e-government initiatives have focused on service delivery and procedural design, neglecting the demand-side dynamics, such as how diverse social groups perceive and use these services (Helbig et al., 2009). The need to balance convenience and security raises further questions about privacy, oversight, and accountability (Bertot et al., 2010). Meanwhile, emerging forms of digital public encounters, including automated services and data-driven decision-making, demand new technical proficiencies and more robust ethical frameworks (Lindgren et al., 2019). Importantly, how governments communicate about digital technologies—and the rhetorical strategies they employ to legitimize new tools—varies significantly across institutional and cultural contexts (Lund-Tønnesen, 2025). This underscores the need to examine not only what digital tools do, but also how they are discursively framed and perceived within specific administrative settings. In all, digital innovations promise a future of interactive governance, and their success hinges on addressing infrastructural gaps and developing trust and legitimacy among stakeholders.

### *AI's Role in Government–Citizen Communication*

While digital tools have long been credited with improving service access and responsiveness (Layne & Lee, 2001), the integration of AI introduces more subtle and discursive changes that merit closer examination. As governments embrace digital transformation, AI is no longer confined to back-end processes and is increasingly embedded in the everyday texture of public interactions. This shift raises new questions about how AI may adapt language and tone to context, thereby reshaping the perceived quality of government communication. Understanding these changes is essential for assessing how AI might influence citizens' trust, satisfaction, and willingness to engage with public institutions.

Prior studies have emphasized digital technologies' potential to reshape government–citizen relationships (Vial, 2019), especially with AI's growing adoption in public sector communications (Van Noordt & Misuraca, 2022; Wirtz et al., 2019; Yigitcanlar et al., 2024). The integration of AI technologies, particularly chatbots and natural language processing systems, has emerged as a promising approach to enhance government–citizen interactions by enabling more responsive and adaptive communication (Androutsopoulou et al., 2019; Pislaru et al., 2024). Researchers have identified multiple benefits of AI implementation in public services, including improved service accessibility, reduced waiting time, and round-the-clock information availability (Larsen & Følstad, 2024; Pislaru et al., 2024). However, using AI in government communication presents multiple considerations and challenges. Research on government chatbots has explored the crucial role of social characteristics and identity design in fostering citizen trust and engagement (Ju et al., 2023; Li & Wang, 2024). Communicative traits, such as emotional intelligence, proactivity, conscientiousness, and professionalism, shape citizen preferences for AI-mediated public communication (Ju et al., 2023; Li & Wang, 2024). Moreover, ethical frameworks and guidelines for AI implementation have been emphasized (Hagendorff, 2020; Jedličková, 2024), particularly with respect to privacy, security, and data-sharing concerns (Campion et al., 2022). Beyond privacy and security, these ethical discussions must be situated within broader critical debates about algorithmic governance.

Scholars have raised concerns that AI-enabled administration may extend bureaucratic rationalization in ways that constrain democratic deliberation and citizen agency (Yeung & Lodge, 2019). Critical AI scholarship warns that algorithmic systems can embed and amplify existing power asymmetries, privileging institutional efficiency over diverse modes of citizen expression (Benjamin, 2019; Noble, 2018). Crawford (2021) further argues that AI systems are not neutral technical artifacts but rather reflect particular political economies and value systems. When applied to government–citizen communication, these critiques suggest that AI modifications—while potentially enhancing "communication quality" by conventional metrics—may simultaneously suppress forms of emotional expression, dissent, or grievance that carry democratic significance. Acknowledging these tensions, the present study proceeds with empirical investigation while remaining attentive to the normative complexities such interventions entail. This requires moving beyond an instrumental understanding of algorithm deployment to examine how both citizens and civil servants actually perceive AI-modified communication.

To understand AI's role in government–citizen communication, a comprehensive evaluation on the citizen side is indispensable. From the citizen's perspective, the successful implementation of AI in public sector communications requires careful attention to citizen preferences and socio-economic factors, including residential environment, employment status, household condition, and education level (Pislaru et al., 2024). Past work indicates that the general public prefers AI to serve in an advisory capacity rather than as an autonomous decision-maker, especially in ideologically charged situations (Haesevoets et al., 2024). In effect, AI has been used more in public service delivery and internal management than in policymaking (Sousa et al., 2019; Van Noordt & Misuraca, 2022). Correspondingly, previous evaluations of the adoption of AI technologies in government–citizen communication have primarily demonstrated improvements in administrative efficiency, with less attention to citizens' perceptions (Larsen & Følstad, 2024). Current trends in government chatbot development are shifting from a focus on information retrieval and service access to citizen consultation and collaboration (Cortés-Cediel et al., 2023). Research on public sector digitalization has demonstrated that citizen feedback plays a central role in shaping how public organizations adapt their digital communication over time (Fossheim & Lund-Tønnesen, 2024). Thus, comprehensive evaluation frameworks are needed that go beyond basic effectiveness metrics to measure public value creation and citizen satisfaction (Cortés-Cediel et al., 2023; Di Vaio et al., 2022).

From the government's perspective, digital transformation requires public communication to carefully craft narratives that address frontline workers' expectations and concerns (Nielsen et al., 2024). Yet, the scholarship on AI and government–citizen communication has undertheorized perceptions of civil servants, overlooking the practical and emotional work they perform in interpreting and responding to public messages. This omission limits our understanding of how communication technologies actually function within administrative processes. Closing this gap requires renewed focus on the internal dynamics of digital governance, particularly how civil servants perceive and engage with emerging tools designed to enhance public interaction.

## Analytical Framework and Hypotheses Development

Despite growing interest in AI's role in public communication, scholarship has yet to provide a comprehensive framework that accounts for both citizen and civil servant perspectives in government-citizen interactions. Most studies have focused either on the citizen side or on internal administrative processes. What remains under-theorized is the communicative exchange between the two, especially when AI intervenes. The present study builds a conceptual foundation that treats communication between the government and citizens as a two-sided, interpretive, and emotionally charged process—one shaped by both relational expectations and cognitive burden. This framework integrates insights from communication studies, behavioral psychology, and public administration, applying them to the distinct realities of public-sector messaging. It recognizes that AI does not merely automate content; it mediates meaning, softens emotion, and shapes how communicative burden and legitimacy are perceived across the exchange.

AI-modified communication, or AI-mediated communication (AI-MC), is a new frontier. It introduces artificial intelligence as a third-party voice that alters how messages are perceived and evaluated. Scholars note that AI-MC involves intelligent systems that modify or generate messages on behalf of humans to achieve communicative goals (Hancock et al., 2020). This differs from traditional computer-mediated communication, where humans operate the tools directly. Instead, AI acts semi-autonomously. Scholars also note that such interventions can restructure interpersonal impressions, leading to more polished but sometimes less authentic interaction (Walther, 1996). Early work on relational behavior in mediated contexts suggests that people adapt to cue-scarce environments by amplifying textual markers of tone and intention (Walther & Burgoon, 1992). More recent studies extend this to algorithmic smart replies and show how AI shapes both perceived cooperation and emotional language, even when users are unaware of its involvement (Hohenstein et al., 2023). This technological mediation forces a reconsideration of how government communication is understood—not only as information exchange but as a designable encounter.

Yet these insights alone cannot explain how citizens and civil servants interpret and act on messages. For citizens, a government reply is more than an answer—it is a cue for satisfaction, legitimacy, and trust. Scholars note that satisfaction arises when expectations are met or exceeded and is shaped by both preexisting beliefs and the actual communicative experience (Oliver, 1980). This idea is extended in the context of technology use, where continued engagement depends on perceived usefulness and the confirmation of prior expectations (Bhattacherjee, 2001). In a public setting, this dynamic becomes more complex. Citizens do not merely evaluate information quality; they assess tone, attention, and respect. When AI modifies a government response, citizens may not be aware, yet they still interpret the message through their expectations. These are social judgments, grounded in expectation-confirmation patterns. Moreover, expectations are not static; expectancy violations—both positive and negative—trigger strong emotional responses that influence perceived appropriateness and credibility (Burgoon, 1993). In digital communication, where tone is harder to judge, these violations are even more salient. A message that feels too cold, too vague, or too polished may signal bureaucratic detachment. However, excessive friendliness might raise suspicion or seem manipulative. Scholars have found that even subtle shifts in tone can violate norms of formality and transparency, altering the entire interpretation of a message (Burgoon & Hale, 1988). Thus, AI-modified content must tread a fine line between softening harsh messages and undermining authenticity. Understanding how citizens process these shifts is vital to assessing communication quality. We acknowledge that framing "communication quality"—defined through dimensions such as clarity, politeness, and perceived empathy—as a desirable outcome reflects a normative stance grounded in administrative efficiency and service-oriented ideals. This framing privileges institutionally legible communication, which may not fully capture the value of emotionally expressive or spontaneous citizen voice. We adopt this framework because it aligns with how public administrators typically evaluate service interactions, while recognizing that critical perspectives—discussed in section 2.2—offer important counterpoints.

Equally important, but often neglected, is how civil servants experience citizen messages. Communication between the government and citizens is rarely symmetrical. Civil servants are required to respond under time pressure, administrative rules, and resource constraints. Scholars note that public interaction with the state is often shaped by administrative burden—defined as the learning, psychological, and compliance costs associated with understanding and addressing public queries (Moynihan et al., 2015). Frontline civil servants assume invisible labor: they must parse emotional language, decode unclear demands, and manage risks. When messages from citizens are emotionally charged or ambiguous, the perceived burden rises. Scholars find that civil servants tend to devalue or resist policies that feel administratively burdensome, even if the policy itself is sound (Burden et al., 2012). In this light, clearer and politer citizen messages—if aided by AI—can ease this burden and improve communication flow. The framework also treats communication as relational. Scholars argue that in computer-mediated communication environments, relationships are not diminished by the absence of nonverbal cues but are instead constructed through verbal density, tone, and timing (Walther, 1992). In communication between the government and citizens, such relational signals manifest in how citizens interpret civil servants' politeness and empathy, and how civil servants perceive citizens' respect or hostility. This links to relational dialectics theory, which sees interaction as a site of tension between control and openness, familiarity and formality (Baxter & Braithwaite, 2008). Scholars claim that dialogue is not just message transmission—it is the co-construction of legitimacy and shared meaning (Baxter, 2004). In government–citizen exchanges, AI can either flatten or enhance these relational dynamics depending on how it is configured.

Another key mechanism is cognitive load, whereby civil servants experience information-processing strain. Scholars have noted that high working memory demands reduce the ability to respond effectively, particularly when messages are complex or poorly structured (Sweller, 1988). Cognitive load theory suggests that simplifying message input, such as by improving clarity and conciseness, enhances responsiveness and decision accuracy (Sweller et al., 1998). If AI could rephrase citizens' inquiries to make them clearer, it would reduce civil servants' information-processing workload. This is not merely a technical matter, as the phrasing of information affects whether civil servants view a citizen's demand as legitimate, actionable, or even worth replying to.

Psychological safety matters in government–citizen communication. Psychological safety—the sense that one can express without fear of dismissal—is key to productive feedback and learning (Edmondson, 1999). In public communication, this means citizens must feel heard, not just answered. When replies convey attentiveness and care, even in the absence of immediate solutions, they foster inclusion. This connects with broader theories of collaborative governance, which emphasize trust-building and dialogic interaction as preconditions for citizen engagement (Ansell & Gash, 2008). In such models, communication is not merely instrumental; it is constitutive of civic legitimacy. AI can assist here by softening tone or expressing acknowledgment; however, it risks being impersonal or generic without strategic design, thereby undermining its relational function.

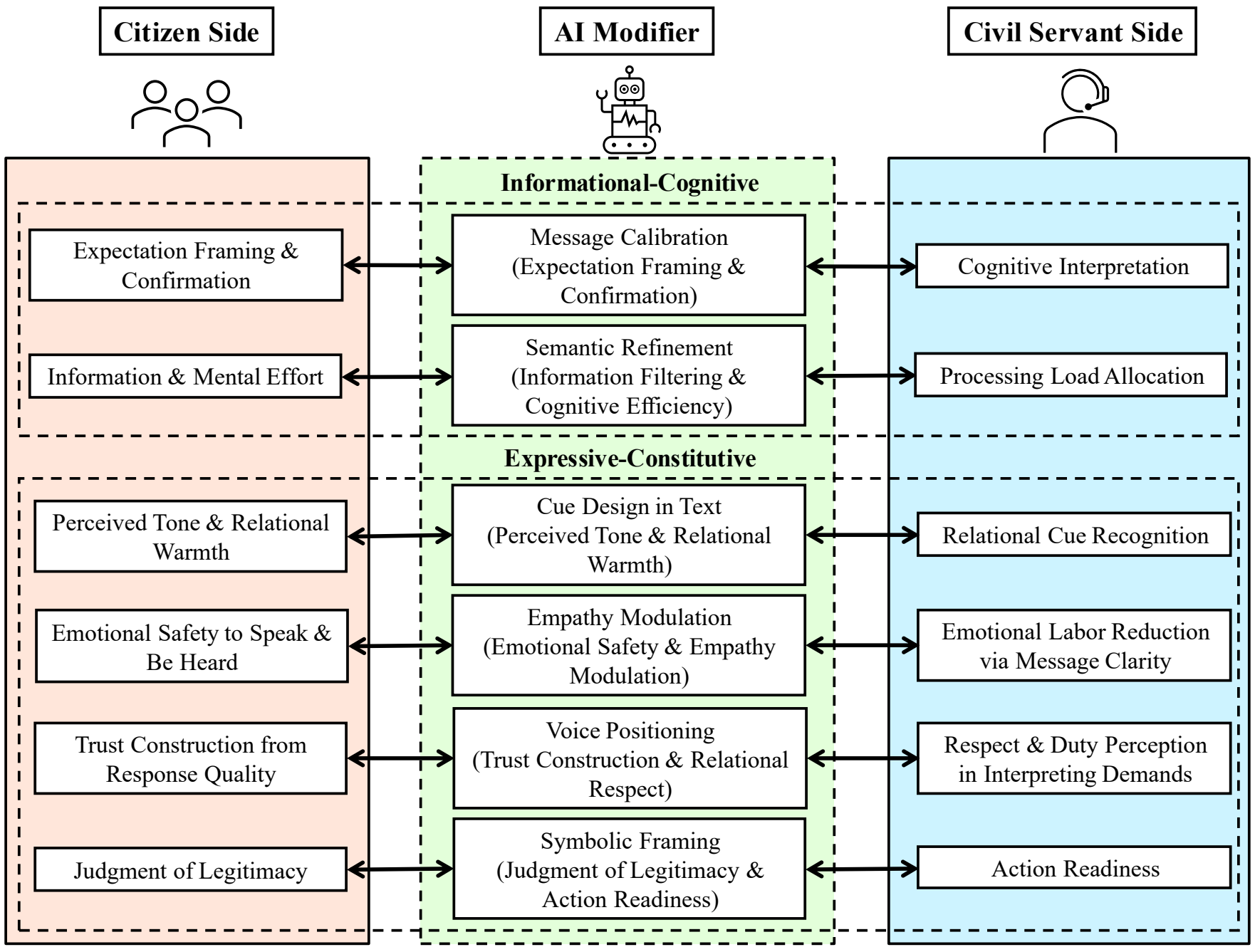

*Figure 1. Analytical framework.*

Empathy is the final but crucial piece. Scholars argue that empathy is not only a personal trait but also a public administrative function (Edlins, 2021). In government–citizen communication, empathic cues from government actors can transform routine replies into moments of connection. It is important to note that AI systems do not experience emotions. Instead, they generate language patterns associated with empathic communication. Throughout this study, references to AI “conveying” or “expressing” empathy should be understood as perceived empathy. These are linguistic features that recipients interpret as empathic, regardless of whether genuine feeling underlies them. Recent studies show that citizens respond more compassionately when made aware of public workers’ burdens, suggesting that empathy is mutual rather than one-sided (Szydlowski et al., 2022). AI systems that neutralize emotional content may inadvertently erase these empathic signals. Alternatively, if trained to detect and amplify appropriate empathy, they may enhance perceived legitimacy. The key is calibration. Neither excessive emotionality nor sterile professionalism will suffice. What matters is fit—between context, content, and emotion.

Together, the literature provides conceptual scaffolding for understanding communication between the government and citizens as a two-sided, mediated, and relational practice. They provide vocabulary for interpreting how citizens and government officials interpret modified messages—in terms of informational quality, affect, tone, burden, and trust. Yet none of these theories were built for AI-assisted communication between the government and citizens. They come from adjacent fields: psychology, public administration, education, organizational theory, and digital interaction. Our study builds the missing bridge. By integrating these perspectives, we show how AI reshapes not just what is said, but also how information is understood, felt, and acted upon—on both sides of the public encounter. The analytical framework is shown in Figure 1.

Before developing specific hypotheses, it is important to recognize that the dimensions of communication quality identified above are not uniform in character. We distinguish between two broad categories: informational-cognitive and expressive-constitutive dimensions. Informational-cognitive dimensions concern the efficient processing and transmission of information content. These include information clarity and ease of comprehension—aspects that reduce cognitive burden for message recipients and facilitate effective information exchange (Sweller, 1988; Sweller et al., 1998). For these dimensions, AI modifications that enhance clarity and structure should yield relatively uncontested benefits, as they align with widely shared goals of reducing misunderstanding and processing friction. Expressive-constitutive dimensions, by contrast, concern how communication expresses voice, performs meaning, and constitutes legitimacy and trust. These include politeness, respectfulness, feelings of being heard, empathic connection, and perceptions of urgency. Unlike informational-cognitive dimensions, expressive-constitutive dimensions are not simply "optimizable." They involve trade-offs between competing values: institutional preferences for standardized, professionally appropriate communication versus democratic values of authentic voice, emotional expression, and communicative agency. Critical AI scholarship warns that algorithmic standardization may enhance surface markers of politeness while suppressing the emotional signals—frustration, urgency, distress—that carry democratic and administrative significance (Benjamin, 2019; Crawford, 2021; Noble, 2018). From this perspective, AI's effects on expressive-constitutive dimensions are inherently contested, as "improvement" reflects particular normative standpoints rather than objective criteria. This distinction structures our hypothesis development. We expect AI modifications to yield consistent positive effects on informationalcognitive dimensions, whereas we anticipate more complex and potentially ambivalent effects on expressive-constitutive dimensions.

Drawing on this integrated framework and the distinction between informational-cognitive and expressive-constitutive dimensions, we develop hypotheses regarding how AI modifications shape perceived communication quality for both citizens and civil servants. From the citizen's perspective, we first consider informational-cognitive dimensions. Cognitive load theory suggests that clear and well-structured messages reduce processing effort and enhance comprehension (Sweller, 1988). Expectation-confirmation theory adds that satisfaction arises when communicative experiences meet or exceed prior expectations (Bhattacherjee, 2001; Oliver, 1980). When government responses are modified to improve clarity and directly address citizen concerns, citizens should experience reduced cognitive effort and higher content satisfaction. These improvements are relatively uncontested, as they serve the shared goal of effective information exchange. Thus:

**H1** Citizens perceive AI-modified government responses as higher in informational-cognitive quality than original responses, in terms of (a) information clarity and (b) response content satisfaction.

For expressive-constitutive dimensions, the picture is more complex, and we distinguish between relational tone and deeper legitimacy construction. Relational communication theory suggests that tone and acknowledgment serve as cues for connection (Baxter, 2004; Walther, 1992). AI modifications that enhance politeness should positively disconfirm citizens' baseline expectations of bureaucratic communication. This effect is relatively direct: AI can readily insert courteous language and soften bureaucratic tone. Thus:

**H2** Citizens perceive AI-modified government responses as higher in expressed politeness than original responses.

However, deeper dimensions of expressive-constitutive quality—feeling heard, empathic satisfaction, and trust—involve more than surface-level tone. These dimensions concern whether citizens perceive genuine institutional responsiveness and whether communication constitutes an authentic connection or merely appears to do so. Critical perspectives caution that AI enhancements may produce formulaic politeness without fostering genuine relational engagement; citizens might perceive AI-modified responses as professionally competent but emotionally hollow. Yet relational communication theory suggests that even mediated warmth can generate perceived connection, and psychological safety theory indicates that feeling heard—even without immediate resolution—fosters trust and continued engagement (Edmondson, 1999; Walther, 1992). These competing theoretical expectations frame the following hypothesis for empirical testing: can AI modifications enhance citizens' deeper relational perceptions, or do such enhancements remain superficial? We assume positive effects, while acknowledging that this assumption is conceptually contested:

**H3** AI-modified government responses elicit more positive citizen relational perceptions than original responses, in terms of (a) feeling of being heard, (b) empathic satisfaction, and (c) trust toward government.

From the civil servant's perspective, we also distinguish between informational-cognitive and expressive-constitutive dimensions. In terms of informational-cognitive dimensions, cognitive load theory suggests that poorly structured or ambiguous messages increase processing strain and impair responsiveness (Sweller, 1988; Sweller et al., 1998). AI modifications that clarify and streamline citizens' messages should reduce this cognitive burden and make responses easier to formulate. These benefits are relatively straightforward, as clearer messages facilitate more efficient administrative processing. Thus:

**H4** Civil servants perceive AI-modified citizen messages as higher in informational-cognitive quality than original messages, in terms of (a) information clarity and (b) easiness of reply.

For expressive-constitutive dimensions, the effects are again more nuanced. Administrative burden scholarship indicates that emotionally charged or hostile messages impose psychological costs on frontline workers (Burden et al., 2012; Moynihan et al., 2015). When AI softens tone and enhances respectfulness, civil servants should experience reduced interpretive and emotional labor, perceiving citizen messages as more polite and respectful. However, this softening involves a trade-off: the same modifications that reduce emotional burden may also strip away communicative signals that civil servants rely on to gauge priority and calibrate their responses. Thus:

**H5** Civil servants perceive AI-modified citizen messages as higher in expressive-constitutive quality than original messages, in terms of (a) expressed politeness and (b) expressed respectfulness. However, the effects of AI modification on urgency and perceived empathy needs illustrate the contested nature of expressive-constitutive dimensions. Critical perspectives on algorithmic communication suggest that when grievances are softened and frustration is tempered, important signals of priority may be muted (Benjamin, 2019; Crawford, 2021). Civil servants often rely on emotional cues—tone, intensity, expressions of distress—to gauge the severity of citizen concerns and allocate attention accordingly. AI's standardization of language may strip away precisely these signals, rendering urgent matters indistinguishable from routine inquiries. This represents the core tension identified by critical AI scholarship: what appears as "improvement" in one metric (e.g., politeness, professionalism) may constitute a loss in another (e.g., communicative authenticity, democratic voice). Similarly, when citizen messages are rendered politer and less emotionally charged, civil servants may perceive less need for empathic engagement—not because the underlying situation is less serious, but because the linguistic markers of distress have been filtered out. Thus:

**H6** AI-modified citizen messages reduce civil servants' perceived (a) information urgency and (b) needed empathy to reply.

## Methodology

We employ a vignette-based survey to collect evaluative assessments from citizens and civil servants regarding government–citizen communication quality. Each participant evaluates both AI-modified and original messages across different topics, with the assignment of AI modification to specific topics randomized across participants. This within-subject design, in which participants serve as their own controls, enhances statistical power and reduces error variance associated with individual differences (Charness et al., 2012; Greenwald, 1976).

The empirical design involves two distinct groups of participants from Mainland China (see Table 1). The citizen group comprises 220 individuals aged 18 or over with prior experience interacting with the government. The civil servant group consists of 214 active civil servants from various government departments. In recent years, Chinese government agencies have actively engaged in online dialogue with the general public through various channels, including official government Weibo accounts, the WeChat government platform, and the Local Leaders Message Board (developed by the People's Daily). The Local Leaders Message Board, in particular, serves as a prominent nationwide platform where citizens submit written complaints, inquiries, and suggestions to local government officials and receive formal written responses—making it well-suited for studying AI's role in text-based government–citizen communication. Research has shown that Chinese local governments are comparably responsive to citizen appeals through these digital channels as similar institutions in democracies (Distelhorst & Hou, 2017). China's administrative system is characterized by hierarchical structures in which local officials face pressure to respond promptly and satisfactorily to citizen inquiries (Chen et al., 2016). At the same time, rapid digitalization has created new demands on local governments to manage high volumes of citizen messages, increasing the practical relevance of AI-assisted communication tools. These features make China an instructive case for examining how AI language modifications affect government–citizen communication, while also raising questions about generalizability to other administrative contexts, which we address in the discussion.

We collaborated with Wenjuanxing, a reputable online survey platform in China, to recruit participants. Participants received monetary compensation via the platform upon completing the survey. Prior to participation, all respondents provided informed consent after reviewing a disclosure statement outlining the study's purpose, procedures, and data usage. Responses were collected anonymously, and no personally identifiable information was retained. The study protocol was reviewed and approved by the Institutional Review Board at the corresponding author's affiliated institution (IRB approval number: HSEARS20241108004). Prior to recruitment, strata were defined based on key demographic characteristics—including age, gender, and education—and participants were randomly selected within each stratum to ensure balanced representation. For civil servants, additional stratification criteria included years of public service experience and frequency of engagement with citizen inquiries, enabling the sample to reflect the heterogeneity of frontline administrative roles. This sampling strategy enhances the internal validity of the study and supports subgroup comparisons across diverse demographic and institutional contexts.

| Variable | Category | Citizens (n=220) | Civil Servants (n=214) |
|---|---|---|---|
| **Gender** | | | |
| | Male | 85 (38.64%) | 104 (48.60%) |
| | Female | 135 (61.36%) | 110 (51.40%) |
| **Age Group** | | | |
| | 18–35 years | 161 (73.18%) | 155 (72.43%) |
| | 36–55 years | 57 (25.91%) | 59 (27.57%) |
| | 56–65 years | 2 (0.91%) | – |
| | 65+ years | – | – |
| **Education Level** | | | |
| | High school / vocational or below | 7 (3.18%) | 1 (0.47%) |
| | Junior college | 19 (8.64%) | 2 (0.93%) |
| | Undergraduate | 175 (79.55%) | 186 (86.92%) |
| | Master's or above | 19 (8.64%) | 25 (11.68%) |
| **Occupation Category** | | | |
| | Civil servant / public institution staff | 18 (8.18%) | – |
| | Enterprise employee | 180 (81.82%) | – |
| | Self-employed | 6 (2.73%) | – |
| | Freelancer | 4 (1.82%) | – |
| | Student | 12 (5.45%) | – |
| | Retired | – | – |
| | Other | – | – |
| **Residential Area Type** | | | |
| | Urban area | 203 (92.27%) | – |
| | Rural area | 17 (7.73%) | – |
| **Interaction Frequency with Government** (past year) | | | |
| | Never | 26 (11.82%) | – |
| | 1–2 times | 102 (46.36%) | – |
| | 3–5 times | 72 (32.73%) | – |
| | 6 or more times | 20 (9.09%) | – |
| **Message Handling Freq.** | | | |
| | Rarely | – | 13 (6.07%) |
| | Occasionally | – | 93 (43.46%) |
| | Regularly | – | 108 (50.47%) |
| **Years of Service** | | | |
| | No more than 1 year | – | 3 (1.40%) |
| | 1–3 years | – | 34 (15.89%) |
| | 4–10 years | – | 146 (68.22%) |
| | Over 10 years | – | – |

*Table 1. Demographic Profile of Respondents*

We categorize citizen messages and civil servant responses into five types of interactive contexts to mirror real-world scenarios: service requests, policy inquiries, complaints, suggestions or feedback, and emergencies or urgent concerns. To ensure a comprehensive and balanced representation, 20 sets of citizen messages and corresponding government responses were meticulously selected from historical interaction records on the Local Leaders Message Board, with four samples per interaction type. This selection process captures the diversity and complexity inherent in government–citizen communications. The citizen messages and government responses then underwent AI modification using a chatbot powered by a large language model, GPT-4o (OpenAI et al., 2024a, 2024b). For citizen messages, the chatbot followed a structured process based on a chain-of-thought breakdown (Wei et al., 2022). This began with identifying the core concern of the message and determining whether it was a complaint, inquiry, or request. The chatbot then softened any emotionally charged or accusatory language to ensure a respectful and polite tone. Clarity was enhanced by rewording complex or ambiguous expressions, ensuring that the message was communicated effectively and concisely. The tone was adjusted to ensure neutrality and constructiveness: aggressive or overly emotional phrases were replaced with polite and straightforward language. Lastly, unnecessary words and repetitive phrases were removed to make the message more succinct while retaining all essential details. For civil servant replies, the chatbot enhanced politeness, empathy, and clarity while maintaining a professional tone. Clarity was improved by simplifying technical or complex language and by breaking down detailed information into more accessible and digestible parts. Where appropriate, reassurance was provided by outlining the steps being taken or setting clear expectations. The list of modification prompts is presented in Appendix A.

The survey integrates a perspective-taking design to engage citizens and civil servants in evaluating communication quality (Batson et al., 1997; Davis, 1983; Galinsky & Moskowitz, 2000). This design fosters a deeper understanding of the communication context, enabling participants to provide more informed and nuanced assessments.

Each citizen participant was randomly assigned six message sets spanning different topics. For each set, participants first read an original citizen message to understand the context, then evaluated the corresponding civil servant response. Across the six sets, three responses were AI-modified and three were original, with the condition-topic pairing randomized across participants to control for topic-specific effects. This generated a total of 1,320 evaluations (220 citizens × 6 responses each). These message sets were presented in random order to eliminate potential order bias. Citizens then evaluated each government response on a 5-point Likert scale across six dimensions: (1) information clarity—ease of understanding the response; (2) response content satisfaction—overall satisfaction with the government's reply; (3) expressed politeness—perceived courteousness of the response; (4) feeling of being heard—the extent to which the response conveyed understanding and addressed the concern; (5) empathic satisfaction—emotional satisfaction with how the response acknowledged the citizen's situation; and (6) trust toward government—the degree to which the response contributed to the participant's trust in government. The trust measure adopted in this study captures the trust-building effect of a specific communicative exchange rather than generalized institutional trust. This operationalization reflects the theoretical understanding that citizen trust in government is not static but develops incrementally through accumulated experiences with public agencies (Porumbescu, 2016; Welch et al., 2005). Because participants evaluated discrete government responses rather than their overall relationship with government, measuring trust contribution is more appropriate than measuring absolute trust levels. This structured evaluation enables a comprehensive assessment of how AI modifications affect citizens' perceptions across multiple dimensions.

Similarly, each civil servant participant evaluated six citizen messages spanning different topics—three AI-modified and three original—with condition-topic assignment randomized across participants. This resulted in 1,284 message evaluations (214 civil servants × 6 responses each). Citizen messages were presented in random order to prevent order effects. Civil servants evaluated these messages on the same 5-point Likert scale across six dimensions: (1) information clarity—how clearly the citizen expressed his/her concern or request; (2) easiness of reply—perceived ease of formulating a clear and constructive response; (3) expressed politeness—perceived courteousness of the message; (4) expressed respectfulness—the extent to which the message conveyed respect for the civil servant's role; (5) information urgency—perceived time-sensitivity of the concern; and (6) needed empathy to reply—the degree of empathic engagement required to respond effectively.

The data analysis is bifurcated into paired t-test and mixed-effect regression analysis. We use a paired t-test to compare the mean ratings of AI-modified responses with those of original responses on the same topic. This test assesses whether the observed differences in ratings are statistically significant to determine the effectiveness of AI modifications. To further elucidate the effects while controlling for potential confounding variables, we employ mixed-effect regression analysis (Gelman & Hill, 2007), which models each dimension of communication quality as the dependent variable; AI modification status serves as the primary independent variable. Control variables include demographic factors: sex, age, and education. For citizens, these also include occupation, type of residence, and frequency of interaction with government agencies. For civil servants, control variables also include frequency of handling public inquiries and years of public service. We also incorporate random effects to account for the nested structure of the data—i.e., the paired responses within each message set (Hox et al., 2017). This approach isolates the effect of AI modifications, ensuring that the observed impacts are attributable to AI rather than to external factors. To enhance the reliability and robustness of the statistical inferences, bootstrapping is employed with 1,000 bootstrap samples.

Bootstrapping is a non-parametric resampling technique that does not rely on strict parametric assumptions, making it particularly suitable for data that may deviate from normality or involve small sample sizes (Davison & Hinkley, 1997). By repeatedly resampling the paired observations with replacement and re-estimating the paired t-test and mixed-effect regression models for each sample, bootstrapping generates empirical confidence intervals for the estimates. This method ensures that the findings are not only statistically significant but also practically reliable, reinforcing the stability and consistency of the results obtained from the primary analyses.

## Results

The results of the paired t-test consistently demonstrate that AI modifications significantly enhance all dimensions of citizens' perceptions of communication quality (see Figure 2). The paired t-test compares the mean ratings for each dimension between AI-modified responses (AId=1) and original civil servant responses (AId=0) across different interaction types: request, inquiry, complaint, suggestion, and emergency. The results indicate significant improvements in all six dimensions when AI modifications are employed. Across all interaction types, AI-modified responses consistently achieved higher satisfaction scores. For example, in the request category, response content satisfaction increased from a mean of 3.54 to 3.84 (95% CI: 3.36–4.00, $p < 0.001$). Similar enhancements were observed in inquiry (3.58 to 3.88, $p < 0.001$), complaint (3.79 to 4.00, $p < 0.001$), suggestion (3.49 to 3.76, $p < 0.001$), and emergency (3.59 to 3.92, $p < 0.001$).

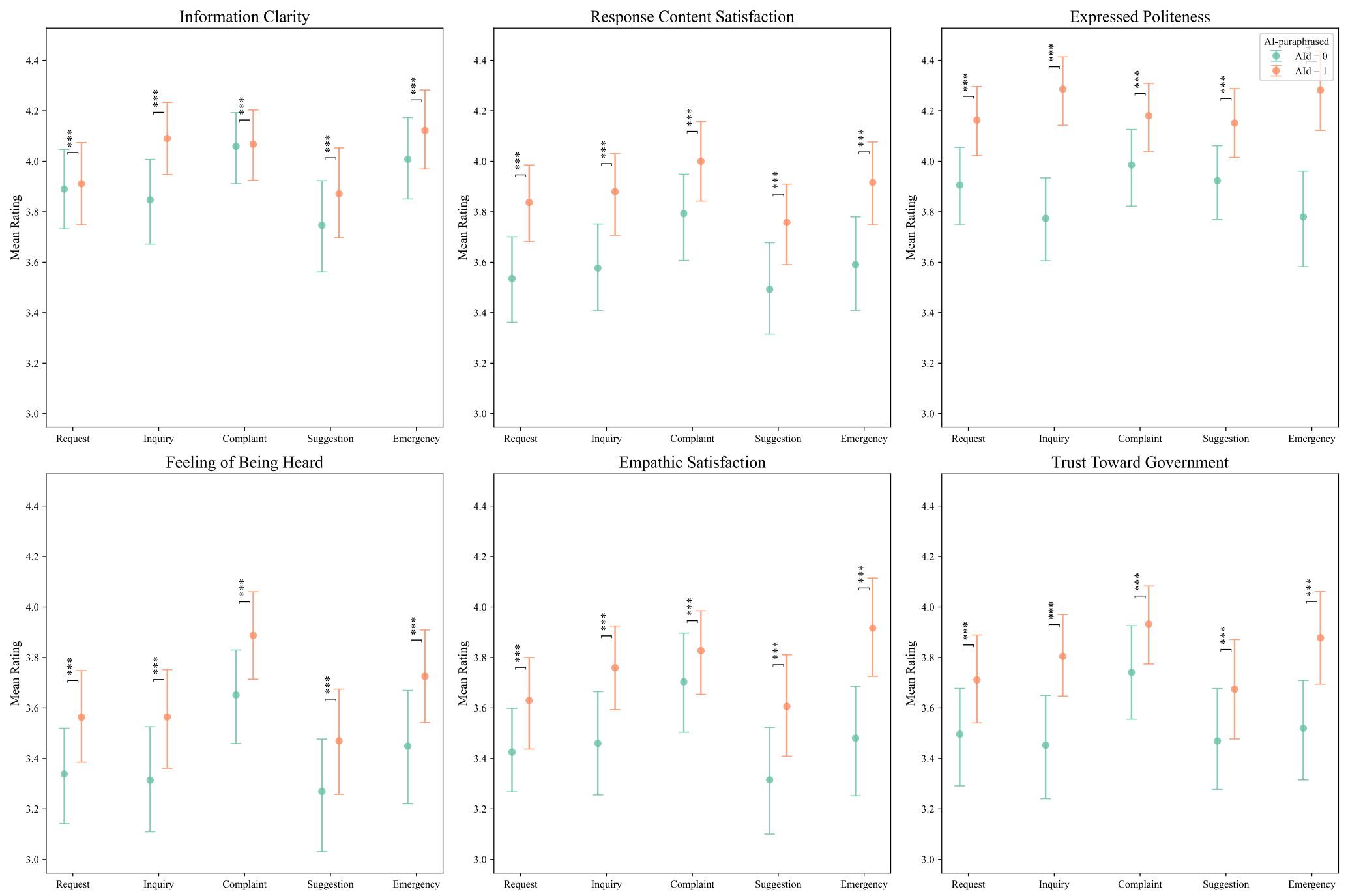

*Figure 2. Comparing citizens' perceptions: paired T-test results.*

The ease with which citizens understood government responses improved notably with AI modifications. While most interaction types experienced significant increases—such as from 3.85 to 4.09 in inquiry ($p < 0.001$)—the request category exhibited a smaller yet significant change from 3.89 to 3.91 ($p < 0.001$). This suggests that AI enhances information clarity, though the extent varies by interaction type. AI interventions significantly elevated perceptions of politeness across all interaction types. In inquiry interactions, expressed politeness ratings rose from 3.77 to 4.29 (95% CI: 4.15–4.41, $p < 0.001$), and in emergency situations from 3.78 to 4.28 ($p < 0.001$). Every topic displayed a statistically significant increase in expressed politeness, underscoring AI's role in fostering respectful communication. AI-assisted responses substantially increased the feeling of being heard. In emergency interactions, the mean rating climbed from 3.45 to 3.73 (95% CI: 3.53–3.91, $p < 0.001$). Similar positive shifts were observed across all other topics, indicating that AI effectively communicates attentiveness and responsiveness to citizen concerns.

AI assistance markedly improved the empathy conveyed through responses: in emergency scenarios, empathic satisfaction ratings rose from 3.48 to 3.92 (95% CI: 3.69–4.11, $p < 0.001$). Trust toward the government was significantly higher after AI modifications. For instance, in inquiry interactions, trust increased from 3.45 to 3.80 ($p < 0.001$). Every interaction type demonstrated a meaningful boost in trust ratings, highlighting AI's potential to enhance the perceived credibility and reliability of the government. All topics showed significant increases in empathic satisfaction scores, suggesting that AI can effectively convey understanding and compassion in communication. Overall, the paired t-test results unequivocally demonstrate that AI-assisted interventions lead to statistically significant improvements in information clarity, response content satisfaction, expressed politeness, feeling of being heard, empathic satisfaction, and trust toward government across various types of government–citizen interactions. These patterns provide preliminary support for H1 through H3, suggesting that AI modifications enhance both informational-cognitive dimensions (clarity, content satisfaction) and expressive-constitutive dimensions (politeness, feeling heard, empathic satisfaction, trust) of citizen-perceived communication quality.

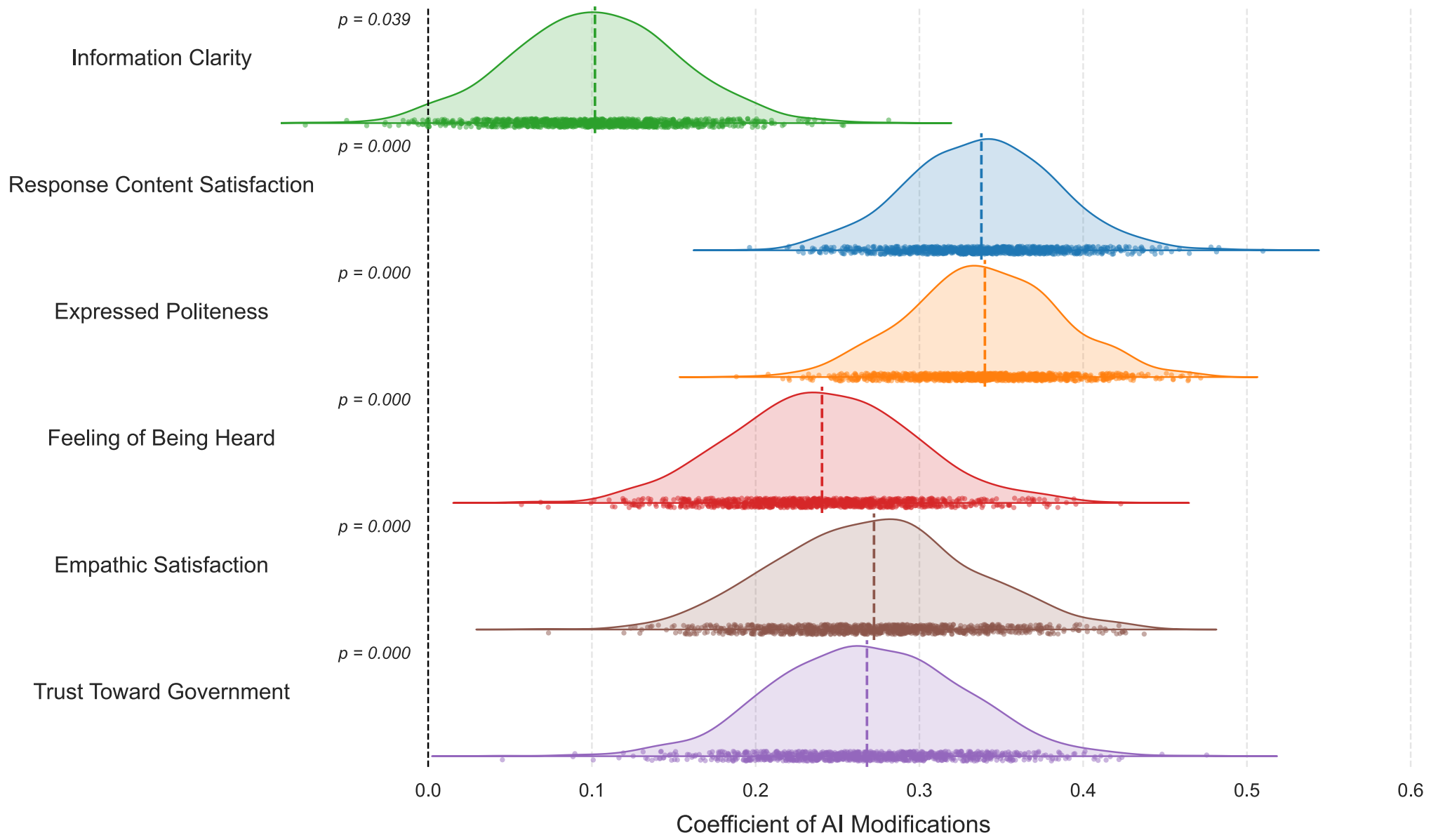

*Figure 3. Ai modifications' impact on citizen perceptions: mixed-effect regression coefficients.*

To further assess the impact of AI modifications while controlling for potential confounding variables such as sex, age, education level, occupation, type of residence, and frequency of interaction with government agencies, we conducted a mixed-effect regression analysis (see Figure 3 and Table 2). This analysis estimates the coefficients representing the effect size of AI modifications on each of the six dimensions. It revealed a mean coefficient of 0.338 (95%

CI: 0.246–0.428, $p < 0.001$), indicating that AI modifications are associated with a substantial increase in satisfaction levels. This reinforces the t-test findings, suggesting that AI plays a significant role in enhancing overall citizen satisfaction with government communications.

With a mean coefficient of 0.340 (95% CI: 0.251–0.429, $p < 0.001$), AI interventions significantly boost perceptions of politeness. This result aligns with the t-test outcomes, emphasizing AI's effectiveness in conveying respectful and courteous communication. The coefficient for information clarity was 0.102 (95% CI: 0.000–0.201, $p = 0.039$). Although the effect size is smaller than for other dimensions, the positive and statistically significant coefficient indicates that AI modifications improve the comprehensibility of communications to citizens. AI-assisted responses had a mean coefficient of 0.241 (95% CI: 0.130–0.355, $p < 0.001$) for the feeling of being heard dimension. This significant positive effect underscores AI's ability to enhance citizens' perceptions of being listened to and acknowledged by government officials. The regression analysis showed a mean coefficient of 0.272 (95% CI: 0.161–0.388, $p < 0.001$) for empathy. This significant positive effect confirms that AI enhancements are effective in conveying empathy, thereby strengthening the emotional connection between citizens and civil servants. AI modifications positively influenced trust, with a mean coefficient of 0.268 (95% CI: 0.152–0.379, $p < 0.001$). This substantial increase indicates that AI can effectively bolster citizens' trust in government responses, which aligns with the observed improvements in satisfaction and politeness. The mixed-effect regression results corroborate the paired t-test findings, demonstrating that AI-assisted interactions have a consistent and significant positive effect across all six dimensions of citizen perceptions. By accounting for various demographic and interaction-related factors, the regression analysis ensures that the observed effects are robust and attributable to the AI modifications rather than external variables. With respect to our hypotheses, the significant positive effects on information clarity and response content satisfaction support H1, confirming that AI enhances informational-cognitive quality. The significant impact on expressed politeness supports H2. Notably, H3—which we framed as theoretically contested—is also supported: feeling of being heard, empathic satisfaction, and trust toward government all showed significant positive effects, suggesting that AI modifications enhance deeper relational perceptions beyond surface-level politeness.

| | Response Content Satisfaction | Expressed Politeness | Information Clarity | Feeling of being Heard | Trust Toward Government | Empathic Satisfaction |
|---|---|---|---|---|---|---|
| **AI modification** | 0.280*** | 0.339*** | 0.105* | 0.244*** | 0.271*** | 0.271*** |
| | (0.0500) | (0.0465) | (0.0499) | (0.0565) | (0.0560) | (0.0573) |
| **Interaction Frequency with Government** (past year) | 0.136*** | 0.0489 | 0.127*** | 0.100** | 0.0694 | 0.0690 |
| | (0.0328) | (0.0306) | (0.0328) | (0.0372) | (0.0368) | (0.0377) |
| **Age** | -0.0279 | -0.0944 | 0.0481 | 0.0448 | 0.00773 | -0.0693 |
| | (0.0563) | (0.0524) | (0.0562) | (0.0637) | (0.0631) | (0.0646) |
| **Education** | -0.0102 | -0.0294 | 0.0217 | -0.0375 | -0.0780 | 0.0122 |
| | (0.0502) | (0.0467) | (0.0502) | (0.0569) | (0.0563) | (0.0576) |
| **Gender** | | | | | | |
| Male | - | - | - | - | - | - |
| Female | 0.124* | 0.0529 | 0.0569 | 0.156** | 0.0746 | 0.0648 |
| | (0.0529) | (0.0493) | (0.0529) | (0.0599) | (0.0593) | (0.0607) |
| **Occupation** | | | | | | |
| Civil servant / public institution staff | - | - | - | - | - | - |
| Enterprise employee | -0.169 | -0.0192 | 0.00986 | -0.152 | -0.185 | -0.0784 |
| | (0.0948) | (0.0882) | (0.0947) | (0.107) | (0.106) | (0.109) |
| Self-employed | 0.124 | -0.143 | 0.361 | -0.00906 | -0.0413 | 0.151 |
| | (0.185) | (0.172) | (0.185) | (0.209) | (0.207) | (0.212) |
| Freelancer | 0.102 | -0.0112 | -0.0976 | -0.0762 | 0.0653 | -0.0365 |
| | (0.212) | (0.198) | (0.212) | (0.240) | (0.238) | (0.244) |
| Student | 0.0631 | -0.0471 | 0.187 | 0.00243 | -0.151 | -0.0389 |
| | (0.148) | (0.137) | (0.147) | (0.167) | (0.165) | (0.169) |
| **Residential Area** | | | | | | |
| Urban area | - | - | - | - | - | - |
| Rural area | -0.143 | -0.299** | -0.257** | -0.191 | -0.149 | -0.161 |
| | (0.0982) | (0.0914) | (0.0981) | (0.111) | (0.110) | (0.113) |
| **Intercept** | 3.404*** | 3.975*** | 3.439*** | 3.259*** | 3.711*** | 3.400*** |
| | (0.254) | (0.225) | (0.242) | (0.292) | (0.279) | (0.290) |
| *N* | 1320 | 1320 | 1320 | 1320 | 1320 | 1320 |

Standard errors in parentheses

* $p < 0.05$, ** $p < 0.01$, *** $p < 0.001$

*Table 2. AI Modifications' Impact on Citizen Perceptions: Mixed-Effect Regression Table*

Understanding how AI-assisted interactions influence civil servants' perceptions and communication effectiveness provides a comprehensive view of AI's role in enhancing overall government–citizen communication. Similar to the citizen analysis, we employed paired t-test to evaluate how AI modifications affect civil servants' understanding of citizens' messages, comparing the mean ratings between AI-modified responses (AId=1) and original responses (AId=0) across the five interaction types: request, inquiry, complaint, suggestion, and emergency. The results reveal significant changes in most dimensions, with varying directions of effect (see Figure 4).

AI modifications enhanced information clarity in most interaction types. For example, in inquiry interactions, perceived message clarity increased from 4.02 to 4.34 (95% CI: 4.23–4.47, $p < 0.001$). Perceived ease of reply exhibited minor yet significant improvements. For instance, in request interactions, easiness of reply increased from 3.70 to 3.84 ($p < 0.001$). We observed similar positive shifts across other message topics, suggesting that AI helps civil servants respond more effectively to citizen demands. Politeness perceptions significantly improved with AI interventions across all interaction types. In complaint interactions, politeness ratings rose from 3.73 to 4.30 ($p < 0.001$), and in emergency situations from 3.33 to 4.15 ($p < 0.001$). These substantial increases highlight AI's effectiveness in fostering polite communication from the civil servant's perspective. AI modifications notably enhanced perceptions of respect. In suggestion interactions, respect ratings increased from 4.09 to 4.33 ($p < 0.001$), and in emergency scenarios from 3.62 to 4.18 ($p < 0.001$). These results indicate that AI helps civil servants feel more respected in their communications. AI's impact on urgency was mixed and generally not significant. In the suggestion ($p >= 0.05$) and emergency ($p >= 0.05$) interactions, the differences in information urgency ratings between AI-modified and original responses were not statistically significant. In some cases, such as requests and inquiries, urgency ratings slightly decreased with AI modifications, suggesting that AI may not effectively convey urgency in these contexts. AI also had a negative effect on needed empathy to reply in several interaction types. In request interactions, empathy ratings decreased from 3.56 to 3.40 ($p < 0.001$), and in complaint scenarios from 3.50 to 3.24 ($p < 0.001$). Conversely, slight improvements were observed in the suggestion (3.31 to 3.39, $p < 0.001$) and emergency (3.30 to 3.40, $p < 0.001$) interactions. These preliminary patterns suggest support for H4 (informational-cognitive quality) and H5 (expressive-constitutive quality in terms of politeness and respectfulness), whereas the mixed findings on urgency and empathy—central to H6—indicate a more complex picture that requires further examination through regression analysis.

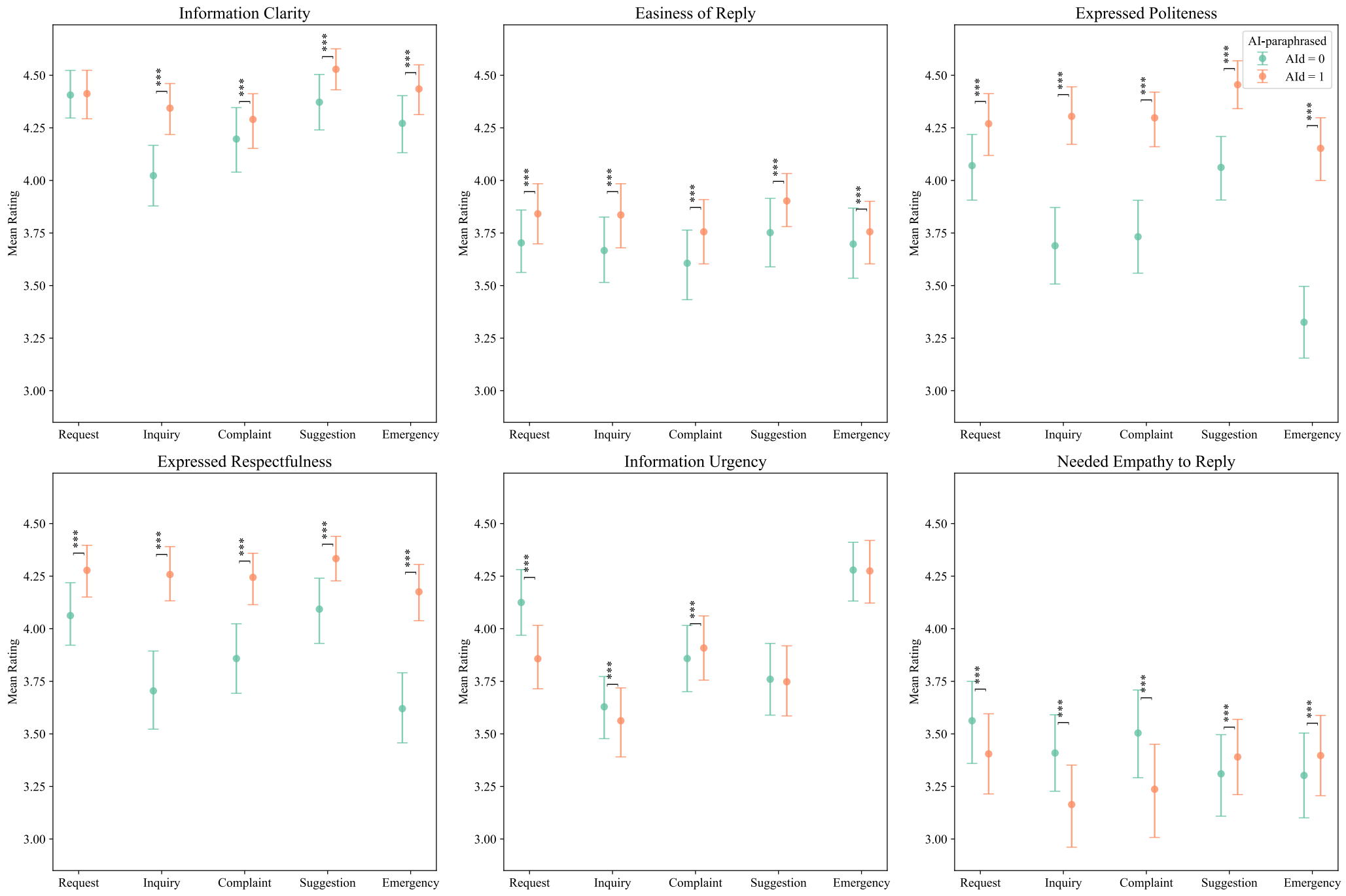

*Figure 4. Comparing civil servant perceptions: paired T-test results.*

The mixed-effect regression analysis for civil servant perceptions assesses the impact of AI modifications while controlling for demographic and interaction-related variables (see Figure 5 and Table 3). The results provide a nuanced understanding of how AI influences each dimension. The regression analysis yielded a mean coefficient of 0.146 (95% CI: 0.065–0.224, $p < 0.001$), indicating that AI modifications increase information clarity in communications.

This suggests that AI helps civil servants understand messages more clearly. The mean coefficient for easiness of reply was 0.131 (95% CI: 0.036–0.227, $p = 0.008$), indicating that AI has a positive and statistically significant impact on civil servants' perceived ease of responsiveness to citizens' demands. This confirms that AI helps make responses timelier and more appropriate. AI interventions had a substantial positive effect on expressed politeness, with a mean coefficient of 0.523 (95% CI: 0.429–0.618, $p < 0.001$). This strong effect underscores AI's significant role in enhancing civil servants' perceived courtesy. AI modifications significantly enhanced expressed respectfulness, with a mean coefficient of 0.390 (95% CI: 0.297–0.476, $p < 0.001$). This result aligns with the t-test findings, highlighting AI's effectiveness in fostering respectful communication. The coefficient for urgency was -0.066 (95% CI: -0.168–0.031, $p = 0.180$), which is not statistically significant. This indicates that AI modifications do not meaningfully affect the perception of urgency in civil servants' responses. The mean coefficient for empathy was -0.110 (95% CI: -0.235–0.007, $p = 0.072$), which approaches but does not reach conventional levels of statistical significance. This suggests a trend toward reduced empathy needed in AI-modified responses, though the effect is not definitively significant. In summary, the civil servant findings support H4: AI modifications significantly enhance informational-cognitive quality, with positive effects on both information clarity and easiness of reply. H5 is also supported, as AI significantly improves expressive-constitutive quality in terms of expressed politeness and respectfulness. However, H6 is not supported. While the coefficients for information urgency and needed empathy are in the hypothesized negative direction—consistent with the critical perspective that AI standardization mutes emotional signals—neither effect is statistically significant. This suggests that concerns about AI systematically dampening urgency and empathy cues may be overstated, or that such effects are context-dependent rather than uniform.

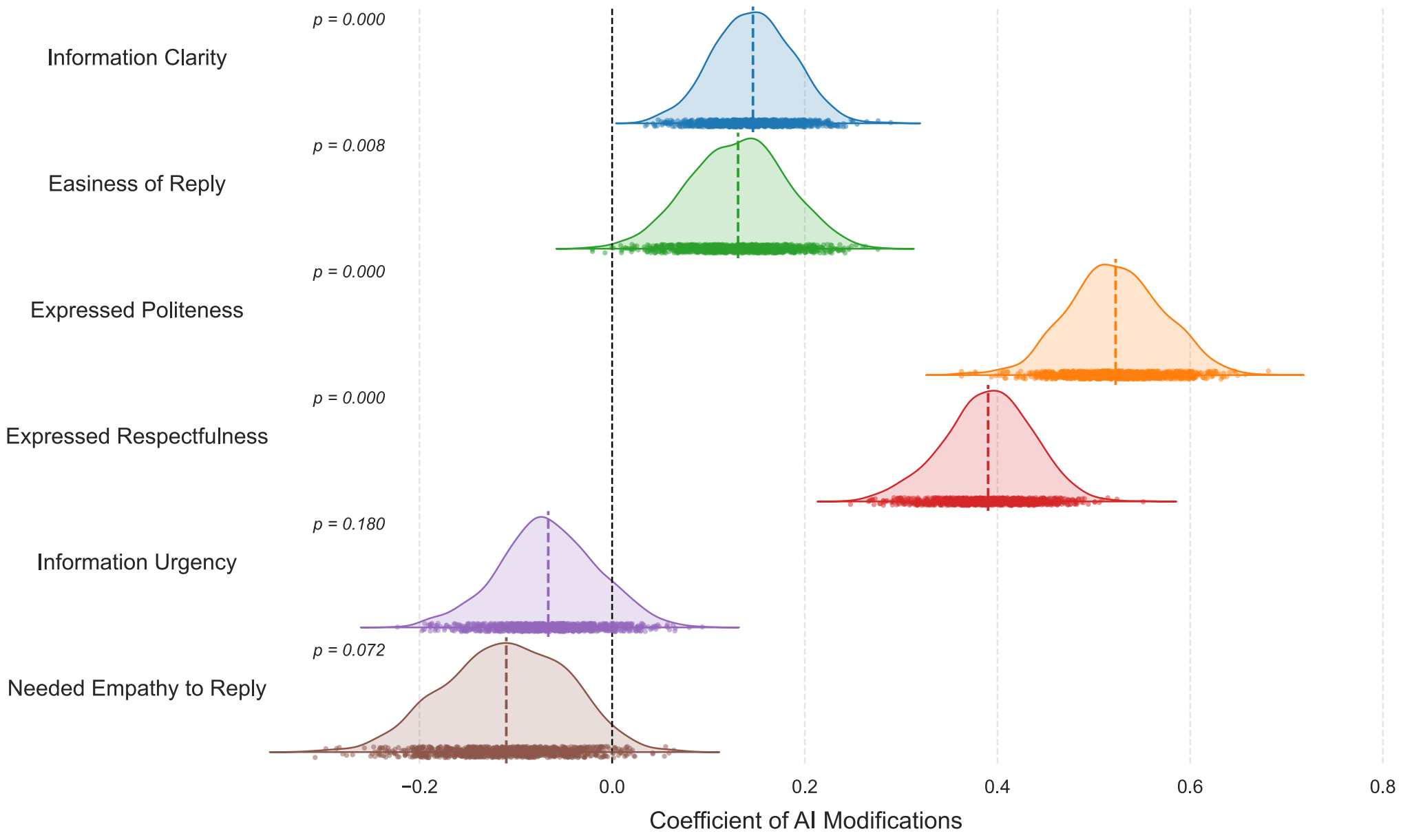

*Figure 5. Ai modifications' impact on civil servant perceptions: mixed-effect regression coefficients.*

## Conclusion and Discussion

This study evaluates how AI language modifications affect the quality of communication between citizens and government under the following interactive contexts: service requests, policy inquiries, complaints, suggestions, and emergency concerns. It empirically establishes that integrating AI into government–citizen interactions could substantially improve communication quality and citizens' perceptions of the government. We focus on key communication quality dimensions from both citizens' and government's sides, including response content satisfaction, expressed politeness, information clarity, feeling of being heard, trust toward government, and empathic satisfaction from the citizens' perspective; and information clarity, expressed politeness, easiness of reply, expressed respectfulness, information urgency, and needed empathy to reply from the civil servants' perspective. Grounded in the evolving landscape of interactive governance and the burgeoning integration of AI into public management processes, this research addresses critical gaps in prior work on the effectiveness and challenges of AI-driven public-sector communication. Specifically, it develops a dual-sided theoretical framework emphasizing how AI simultaneously shapes citizens' expectations and reduces civil servants' burden, bridging previous theories from communication, psychology, and public administration.

The paired t-test results demonstrate significant improvements across the six dimensions of citizen perception of communication quality following the adoption of AI modifications in the written language communication process. In particular, satisfaction with response content increased notably across all five interaction contexts. Similarly, information clarity, expressed politeness, feeling of being heard, empathic satisfaction, and trust toward government all exhibited substantial enhancements. For citizens, aligned with our theoretical framework on expectation-confirmation and relational communication, AI yields clearer and politer responses that enhance citizens' expectations and relational warmth, thereby fostering trust toward government institutions. These results are consistent with prior research suggesting that AI leads to more respectful and considerate communication (Ju et al., 2023; Li & Wang, 2024). Moreover, the significant improvements in information clarity and citizens' feeling of being heard align with the notion that AI can clarify complex information and demonstrate attentiveness to citizen concerns (Cortés-Cediel et al., 2023; Pislaru et al., 2024). The significant improvements in citizens' feelings of being heard and empathic satisfaction reinforce the notion that AI modifications effectively acknowledge and address citizens' concerns, thereby fostering a sense of understanding. The findings also demonstrate consistent and significant increases in trust and satisfaction with government responsiveness, underscoring AI's role in strengthening emotional connections between citizens and civil servants. Prior studies highlighted the complexity of building and maintaining positive relationships between citizens and government institutions through the internet (McNeal et al., 2008; Vial, 2019). Communication barriers inherent in digital government platforms, such as bureaucratic tone, ambiguous phrasing, or perceived indifference, can erode trust. Our findings indicate that AI-assisted interactions may help address some of these challenges: enhancing information clarity and expressed politeness to reduce misunderstandings, fostering a stronger feeling of being heard, and conveying empathy and emotional attentiveness. The mixed-effect regression analysis further substantiates these findings. The regression results underscore AI's effectiveness in enhancing key dimensions of communication quality, even after controlling for demographic and interaction-related variables. These patterns reflect a deeper theoretical architecture. As our conceptual framework highlights, AI-mediated communication reconfigures not only information clarity but also relational expectations, emotional cues, and interpretive burden on both citizens and civil servants. It does not simply mean that messages are more explicit and polite; rather, it further signifies that AI recalibrates how legitimacy, empathy, and trust are constructed across the exchange.

| | Information Clarity | Expressed Politeness | Easiness to Reply | Expressed Respectfulness | Information Urgency | Needed Empathy to Reply |
|---|---|---|---|---|---|---|
| **AI modification** | 0.146*** | 0.519*** | 0.133** | 0.390*** | -0.0678 | -0.113 |
| | (0.0408) | (0.0490) | (0.0489) | (0.0456) | (0.0504) | (0.0628) |
| **Message Handling Freq.** | -0.00185 | 0.0557 | 0.0687 | 0.0177 | -0.0187 | -0.168** |
| | (0.0341) | (0.0410) | (0.0409) | (0.0381) | (0.0422) | (0.0524) |
| **Years of Service** | 0.154*** | 0.0966* | 0.124** | 0.131** | 0.109* | 0.206*** |
| | (0.0391) | (0.0470) | (0.0469) | (0.0437) | (0.0483) | (0.0601) |
| **Gender** | | | | | | |
| Male | - | - | - | - | - | - |
| Female | 0.0217 | -0.0325 | -0.0646 | -0.00583 | -0.0367 | 0.138* |
| | (0.0415) | (0.0499) | (0.0498) | (0.0464) | (0.0513) | (0.0638) |
| **Age** | -0.104* | 0.0465 | 0.0854 | -0.105 | -0.0522 | 0.204* |
| | (0.0527) | (0.0632) | (0.0631) | (0.0588) | (0.0650) | (0.0809) |
| **Education** | 0.0229 | -0.224*** | -0.0283 | -0.149* | -0.0409 | 0.0775 |
| | (0.0562) | (0.0675) | (0.0674) | (0.0628) | (0.0694) | (0.0864) |
| **Intercept** | 3.855*** | 4.010*** | 3.161*** | 4.038*** | 3.868*** | 2.652*** |
| | (0.222) | (0.270) | (0.265) | (0.250) | (0.278) | (0.339) |
| *N* | 1284 | 1284 | 1284 | 1284 | 1284 | 1284 |

Standard errors in parentheses

* $p < 0.05$, ** $p < 0.01$, *** $p < 0.001$

Table 3. AI Modifications' Impact on Civil Servant Perceptions: Mixed-Effect Regression Table

For civil servants, AI helps transmit clearer and politer messages from citizens, which makes civil servants feel respected and anticipate less burden to respond. This aligns directly with our theoretical integration of cognitive load and administrative burden, where AI modification reduces civil servants' emotional labor and cognitive effort by providing well-structured and respectful communication. These enhancements suggest that AI modification tools help civil servants communicate more effectively and courteously, thus reducing misunderstandings and fostering a more respectful interactive environment. This aligns with the findings of Larsen and Følstad (2024), who noted that AI can enhance the professionalism and courtesy of public service communication. However, our results also reveal that the impact of AI on information urgency and the need for empathy to reply is context-specific. Requests and policy inquiry messages became less urgent after AI rewording; urgency ratings in complaints, suggestions, and especially emergency messages remained stable, indicating that AI can preserve perceived urgency when the underlying content strongly conveys it. AI lowered civil servants' perceived need for empathic response in requests, inquiries, and complaints, which may reflect the AI's neutralization of emotional tone in contentious or routine exchanges. Yet, in suggestions and emergencies, AI modifications slightly increased the perceived need for empathy, suggesting that the AI may enhance emotional tone in collaborative or high-stakes contexts. Overall, our empirical results demonstrate that in more routine or negatively framed interactions (e.g., complaints), the AI rewording strips away the strongly emotive language that would otherwise signal a need for high empathy. In contrast, for suggestions and emergencies in which the AI may insert polite expressions of acknowledgment or concern, the reworded messages prompt civil servants to perceive a higher emotional dimension.

The logic under which AI shapes messages—whether through amplification or reduction of emotional intensity—depends heavily on underlying design instructions and ethical considerations established by public administrators (Campion et al., 2022; Hagendorff, 2020; Wirtz et al., 2019). Thus, the practical issue at hand is how governments choose to calibrate AI's emotional amplification or reduction. Critically, it reminds us to reflect on whose conception of "effective communication" guides these choices. Citizens may value emotional authenticity and the ability to express frustration, while civil servants may prioritize clarity and manageable workloads; what serves one party's interests may not serve the other's. Public administrators must consider whether AI will be designed to isolate civil servants' empathic reactions to ensure consistent, objective, and fair administrative responses, thereby safeguarding justice from emotional biases. Alternatively, should AI deliberately amplify emotional urgency—such as emphasizing a citizen's distress or tragedy—to stimulate stronger empathetic responses from civil servants, thereby fostering deeper engagement and improved responsiveness? These are not merely technical questions but normative ones that require deliberation about the values embedded in communication design. Previous research has noted challenges in integrating emotional intelligence into public sector communication (Ju et al., 2023; Li & Wang, 2024). We argue that governments must carefully deliberate on how AI instructions might enhance or attenuate emotional content, remaining attentive to whose interests such configurations serve. Effective government–citizen interaction requires frameworks that balance efficiency with empathy, professionalism with sensitivity, and responsiveness with impartiality (Edlins, 2021; Mussagulova, 2026). Thus, our findings not only identify the complexities of harnessing AI's emotional capabilities but also encourage policymakers and researchers to thoughtfully design AI instructions that strategically leverage emotional amplifiers and reducers. However, such design choices carry risks: configurations that benefit administrative efficiency may not equally serve citizen voice, and vice versa. A related consideration is communicative authenticity. When AI modifies messages, questions arise about whose voice is ultimately conveyed. Citizens may wonder whether their original intent and emotional expression remain intact after algorithmic revision; civil servants may reflect on whether AI-enhanced responses genuinely represent their professional judgment and interpersonal engagement. These concerns do not negate AI's potential benefits but highlight the need for thoughtful implementation that balances communicative efficiency with respect for authentic self-expression.

To better understand why these tensions arise, the theoretical foundation of AI as computational agents that act given inputs (percepts) to achieve the best expected outcome offers a lens for understanding both the successes and limitations of AI in government–citizen communications (Hancock et al., 2020; Russell & Norvig, 2010). AI systems excel when performance goals—such as clarity and politeness—can be explicitly defined and quantitatively measured, enabling the agent to systematically optimize its responses. However, this same rationality highlights why AI struggles with tasks like conveying urgency or empathy, where success depends on tacit, context-sensitive judgments that resist formal codification. Rather than viewing AI as a universal communication optimizer, we might instead conceptualize it as a contextual mediator capable of strategic emotional modulation depending on the specific goals of government–citizen interactions. To this point, we have sought to connect technology to broader e-government. The early digital governance paradigm often envisioned a wholesale technological replacement of traditional administrative processes (Dunleavy et al., 2006; Ho, 2002)—a view that implicitly assumed that all governance functions could be effectively parameterized, optimized, and digitally executed. Our findings suggest a new perspective on technological determinism by illustrating how AI-mediated communication embodies both remarkable capabilities and inherent limitations. Just as AI rational agents excel with welldefined parameters but struggle with contextual emotional intelligence, government–citizen interactions have similarly demonstrated greater effectiveness when implemented strategically rather than universally. This reinforces the importance of a dual-sided, relationally sensitive framework. AI systems do not simply process messages. They transform how communicative burden and legitimacy are distributed across actors. As we propose, effective communication between the government and citizens depends on understanding the interpretive experiences of both citizens and civil servants, and how AI interacts with those experiences across cognitive, emotional, and symbolic levels. Yet this transformation is not inherently neutral—it raises important questions about what is gained and what may be lost when AI mediates public communication.

Our findings, viewed through the lens of critical AI scholarship, surface these tensions. While our results demonstrate measurable improvements in clarity and politeness, AI's effects on perceived urgency and the need for empathic response were inconsistent—improving in some contexts while remaining unchanged or diminishing in others. This pattern suggests that AI may function not merely as a communication enhancer but as a form of discursive standardization that recalibrates citizen expression toward institutionally preferred norms (Benjamin, 2019; Noble, 2018). When emotionally charged language is softened and grievances are rendered "polite," AI risks muting the communicative signals—frustration, urgency, distress—that frontline workers rely on to prioritize cases and allocate resources appropriately. Yet our findings also complicate a purely critical reading: AI modifications increased perceived empathy needs in emergencies and suggestions, indicating that the technology does not uniformly flatten emotional content but rather modulates it in context-dependent ways. This nuance suggests that the risks of AI mediation are not inevitable but contingent on design choices. Following Yeung and Lodge (2019), we argue that algorithmic tools in public communication demand deliberate calibration—not to maximize efficiency alone, but to preserve space for citizen voice and the relational dynamics that underpin administrative legitimacy and public trust. The challenge for public administrators is thus not whether to deploy AI, but how to configure it in ways that enhance communicative quality without obscuring the substantive concerns embedded in citizen expression.

We acknowledge that the study suffers from limitations in empirical design. Our AI modifications primarily focused on written language, such as politeness and clarity, which may not capture all facets of effective communication, such as cultural sensitivity or context-specific nuances. The mixed results regarding information urgency and needed empathy indicate areas for further exploration. Additionally, our study was conducted in the Chinese administrative context, where hierarchical communication norms and relatively formal

government–citizen exchanges may shape expectations for both message tone and government responsiveness. While we expect that AI's effects on clarity and politeness may generalize across governance systems, the magnitude of these effects may differ in contexts with less formal communication norms or different expectations for government responsiveness. Researchers should exercise caution when extending these findings to other administrative cultures. A further limitation concerns ecological validity. Participants in our study evaluated messages written by others rather than their own communications, which may not fully capture how individuals perceive AI modifications to their personal expressions. Moreover, participants were unaware that some messages had been AI-modified. In real-world implementation, awareness of AI mediation could alter perceptions: citizens might feel their authentic voice was filtered, or civil servants might view AI-enhanced responses as less genuinely empathic. Future research should examine whether disclosure of AI involvement affects perceptions of communicative sincerity and trust. Future studies could investigate how different AI design features influence these dimensions and explore the potential for integrating more advanced emotional intelligence capabilities into AI systems. Longitudinal studies could also assess the long-term effects of AI-assisted communication on citizen trust and engagement, providing a deeper understanding of how sustained AI interactions influence public perceptions and relationships with government institutions.

In conclusion, this study underscores the significant potential of AI-assisted interactions to enhance the quality of communication between citizens and the public sector. By improving key perceptual dimensions—such as satisfaction and trust—AI can play a pivotal role in fostering more positive and trusting relationships between citizens and government institutions. However, the challenges identified in conveying urgency and empathy highlight the need for ongoing refinement and thoughtful implementation of AI technologies. As governments continue to embrace digital transformation, AI tools must be designed and deployed ethically and effectively in order to achieve the desired improvements in public service communications and maintain citizens' trust and satisfaction.